\documentclass[a4paper,11pt]{article}

\usepackage{pos}
\usepackage{lipsum} 

\usepackage{xspace}

\newcommand{\xmax}{\ensuremath{X_{\rm max}}\xspace}

\newcommand{\shrink}{\vspace{-0.1cm}}

\usepackage{lineno}

\usepackage{cleveref}
\usepackage{wrapfig}
\usepackage{subfiles} 

\crefformat{figure}{#2Fig.~#1#3}
\crefformat{equation}{#2Eq.~#1#3}
\crefformat{table}{#2Tab.~#1#3}

\title{The Surface Array of IceCube-Gen2}

\ShortTitle{The Surface Array of IceCube-Gen2}

\author{The IceCube-Gen2 Collaboration \\{\normalsize \normalfont(a complete list of authors can be found at the end of the proceedings)}\\}

\emailAdd{alan.coleman@physics.uu.se}
\abstract{

The science goals of IceCube-Gen2 include multi-messenger astronomy, astroparticle and particle physics. To this end, the observatory will include several detection methods including a surface array and in-ice optical sensors. The array will have an approximately 8\,km$^2$ surface coverage consisting of elevated scintillator panels and radio antennas to detect air showers in the energy range of several 100\,TeV to a few EeV. The observatory’s design is unique in that the measurements using the surface array can be combined with the observations of $\geq$\,300\,GeV muons, produced in the hadronic cascades, using the optical detectors in the ice. This allows for an enhanced ability to study cosmic-ray and hadronic physics as well as to boost the sensitivity for astrophysical neutrinos from the southern sky by reducing the primary background, atmospheric muons. We will present the baseline design of the surface array and highlight the expected scientific capabilities.

\vspace{4mm}
{\bfseries Corresponding author:}
Alan Coleman$^{1,2*}$\\
{$^{1}$ \itshape Dept. of Physics and Astronomy, Uppsala University, Box 516, S-75120 Uppsala, Sweden}\\
{$^{2}$ \itshape Bartol Research Institute and University of Delaware, Dept. of Physics \& Astronomy, Newark, DE, USA}\\[4mm]
$^*$ Presenter

\ConferenceLogo{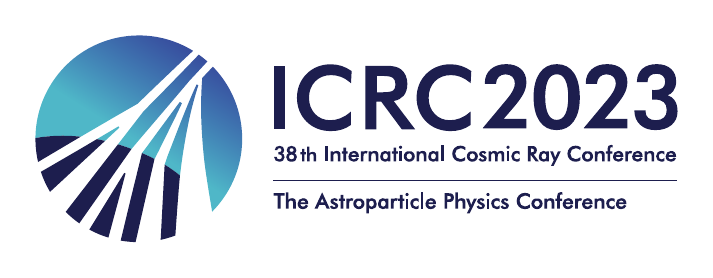}

\FullConference{The 38th International Cosmic Ray Conference (ICRC2023)\\ 26 July -- 3 August, 2023\\ Nagoya, Japan}
}

\begin{document}

\maketitle

\section{Science goals and motivation}
\begin{wrapfigure}{r}{0.5\textwidth}
  \shrink
  \begin{center}
    \includegraphics[width=0.499\textwidth]{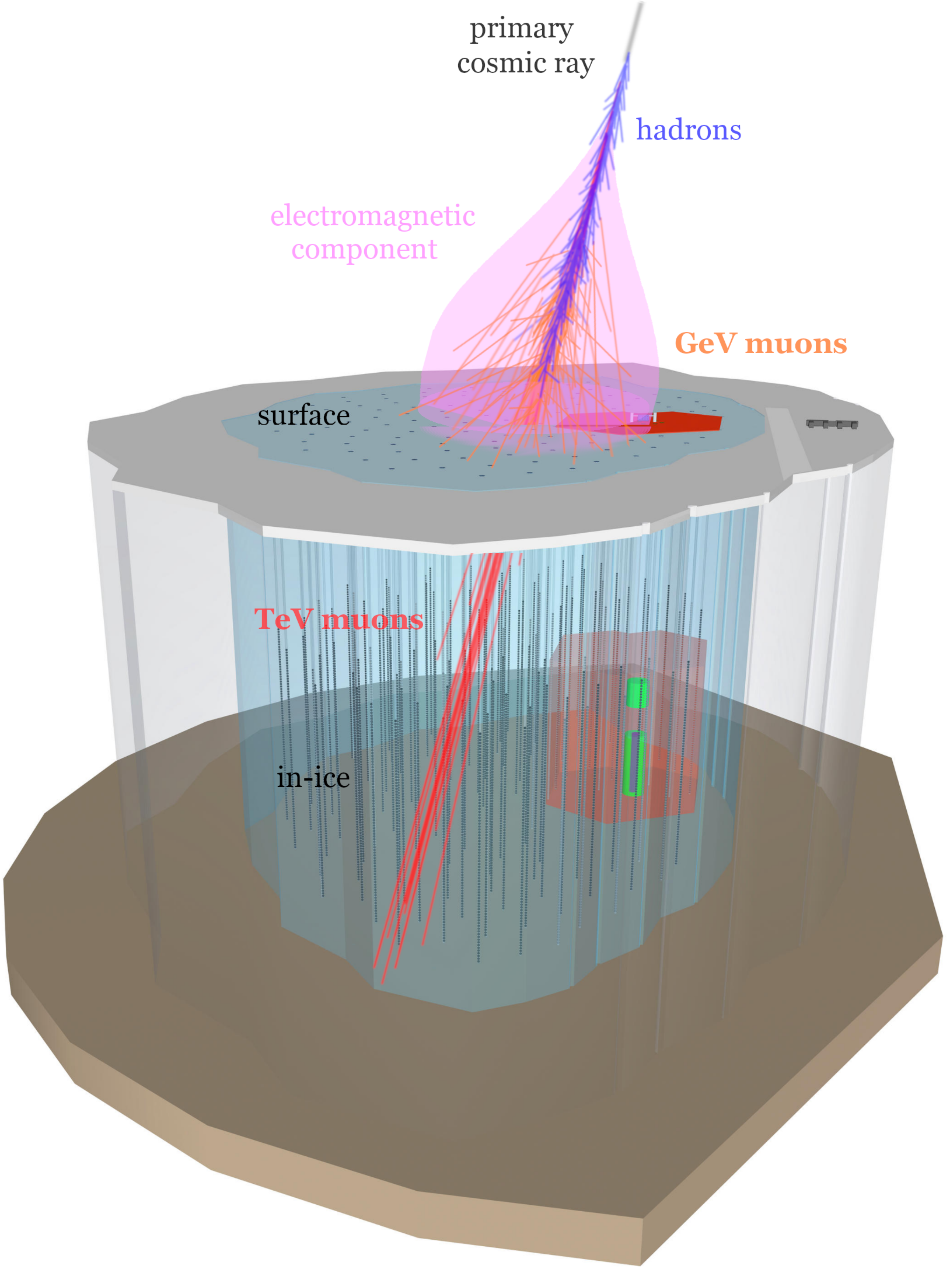}
  \end{center}
  \shrink
  \caption{Schematic of an air shower being observed at IceCube-Gen2. The current observatory is highlighted in red while the future size is shown in blue.}
    \label{fig:Gen2Schematic}
    \shrink
\end{wrapfigure}
IceCube-Gen2~\cite{ishihara:2023icrc} is the proposed expansion to the existing IceCube Observatory~\cite{Aartsen:2016nxy}, located at the South Pole. The observatory has already been the source of several key measurements in both astroparticle physics and particle physics, enhancing our understanding in both fields. The current observatory includes an in-ice array of optical modules, burred 1.5--2.5\,km below the surface of the ice, and covering 1\,km$^3$. The surface array, IceTop, is comprised of water Cherenkov tanks, covering the 1\,km$^2$ surface above the in-ice array~\cite{IceCube:2012nn}. Both of these components of the observatory will be enhanced in IceCube-Gen2, see Ref.~\cite{ishihara:2023icrc}.

The inclusion of the surface array in the next generation of the observatory is motivated by the previous successes and the critical role that the surface array has for improving the understanding of air-shower physics. The unique design, wherein the low-energy particles ($\sim$\,GeV muons and EM particles) can be detected on the surface while $\gtrsim$\,300\,GeV muons can be detected in the ice (see~\cref{fig:Gen2Schematic}), makes the observatory a leading experiment for studying particle physics via air showers.

Indeed, many of the cosmic-ray science goals of IceCube-Gen2 are derived from leveraging this type of coincident information, including a determination of the mass of individual primary nuclei. The evolution of different mass groups as a function of energy and/or direction is important for understanding the transition from galactic to extra-galactic sources. Further, a better understanding of hadronic physics at these energy scales will enhance the measurements for all past and future experiments, and thus for the field as a whole.

The surface array also aids in the neutrino science goals by better characterizing the background for in-ice studies. This background, $\gtrsim$\,300\,GeV muons, is a result of the high-energy processes in air showers and is orders of magnitude larger than all other backgrounds. The expected rate depends on the energy spectra of individual mass groups, along with the typical production of high-energy muons during the shower production. The ability of a single observatory to be used to study its own primary background will be enhanced by the expanded detection capabilities of IceCube-Gen2. The surface array can also provide a veto for this type of background by directly tagging events as being due to an air shower based on the surface activity.

In this proceeding, we will elaborate on the design of the surface component of IceCube-Gen2 and quantify the ability with which the surface array will be able to achieve these science goals.

\begin{figure}[tbh]
  \shrink
  \begin{center}
    \includegraphics[width=0.95\textwidth]{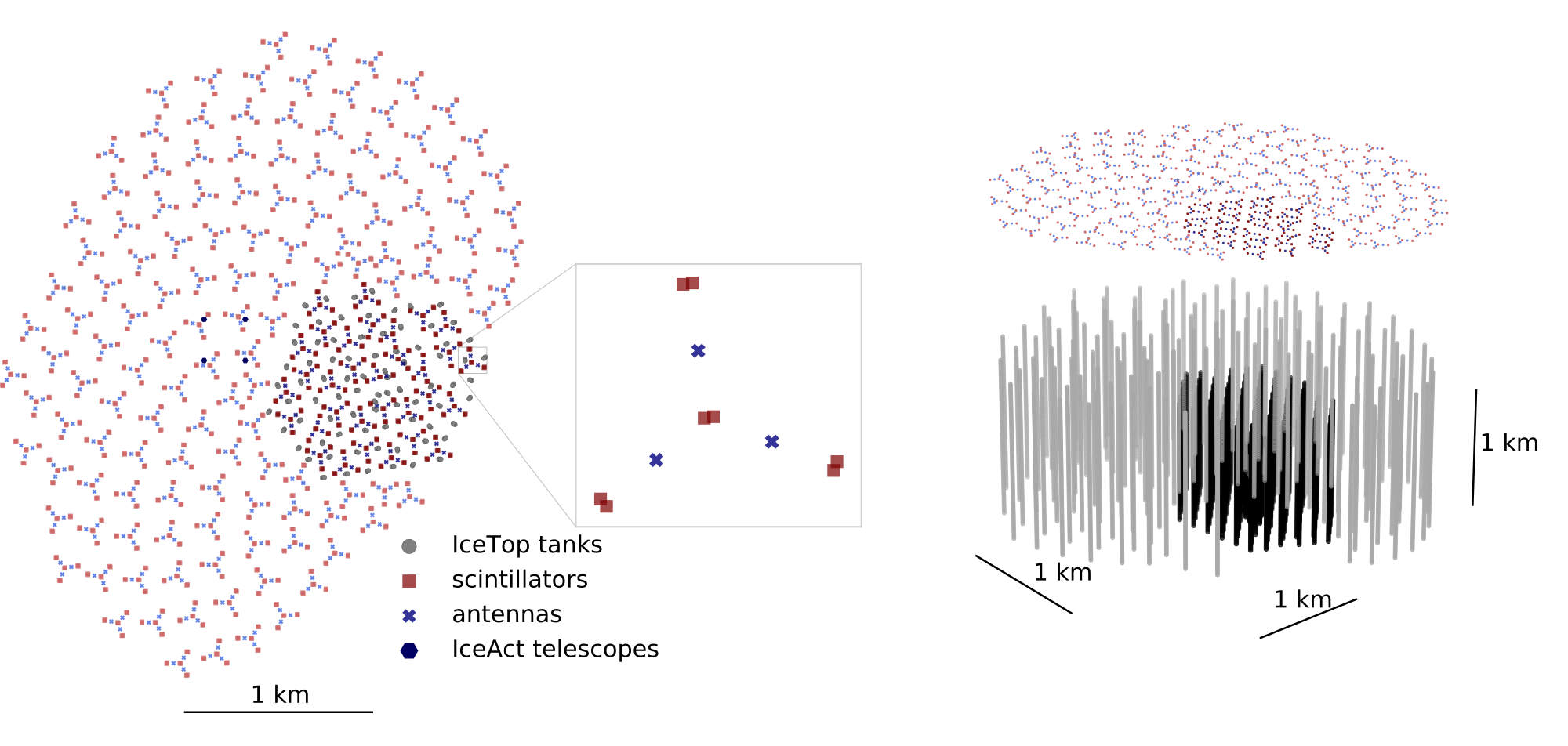}
  \end{center}
  \shrink
  \caption{Schematic of the layout of IceCube-Gen2. The left figure is the overhead view of the surface instrumentation, with the inset highlighting the layout of a surface station. The right figure is the 3D representation, with each black dot indicating an in-ice detector.}
    \label{fig:ObservatoryDiagram}
\end{figure}
\section{Design and air shower detection}

The IceCube-Gen2 surface array is planned to have stations comprised of elevated scintillator panels and radio antennas, see~\cref{fig:ObservatoryDiagram}. The station design is based on the planned Surface Array Extension~\cite{Schroder:2019suq} and will include eight panels, grouped in pairs, with a triangular layout. A radio antenna will be placed approximately halfway down each of the three arms. The stations will be placed above each of the in-ice optical strings and will thus have a typical inter-station-spacing of about 250\,m. In the center of the array will be four IceAct stations, separated by about 200\,m, that comprise seven Cherenkov telescopes (see Ref.~\cite{paul:2023icrc}).

Located at the center of the station will be a \emph{field-hub} where both the surface and in-ice detectors for that station/string will be read out and digitized. The triggering system for the observatory is planned to be global, where each detector component can initiate the readout of the entire observatory. This method allows for flexible triggers that can target specific science goals while not losing any potentially useful information. For air shower detection, it is envisioned that events will be triggered via the scintillator panels or ice-Cherenkov tanks. However, for inclined air showers that largely miss the surface array but have high-energy muons that are detected in the ice, readout of the surface array will also be initiated by the in-ice triggers. This is important for providing a veto for $\nu_{\mu}$ studies and is discussed more below. In all cases, the antennas will be externally triggered.

This design constitutes an increase in instrumented area and volume by about a factor of eight for the surface and in-ice arrays, respectively. However, the increase in aperture for coincident events, i.e. those that will be most useful for mass composition and hadronic physics studies, will increase by a factor of about thirty. Further, the phase-space for studying $\gtrsim$\,300\,GeV muon production will also increase, as the maximum zenith angle for coincident detection will be $\simeq\,68^\circ$ as opposed to the current limit of $\simeq\,39^\circ$.

The capabilities of the surface array for air shower detection is shown in the left panel of~\cref{fig:xmax_and_count}. This plot shows the number of expected events as a function of energy based on the cosmic ray flux model H4a~\cite{Gaisser:2011klf}. The array should see about 100 events at-and-above the ankle ($\sim$\,$10^{18.5}$\,eV) each year. After a single year of operation, the energy range over which the cosmic rays are primarily of galactic origin can be studied. The hashed lines indicate the number of events that will also have high-energy muons in the optical detector, according to the Sibyll-2.3d interaction model~\cite{Riehn:2019jet}. After ten years of operation, measurements of the primary mass can be made up to $10^{19}$\,eV, covering several observed mass-transitions and providing important cross-checks across two order of magnitude in energy with the measurements of the Pierre Auger Observatory and Telescope Array~\cite{Yushkov:2020nhr,TelescopeArray:2018xyi}.

\section{Improved understanding of hadronic and cosmic-ray physics}
\begin{figure}[tb]
  \begin{center}
  \shrink
    \includegraphics[width=0.445\textwidth]{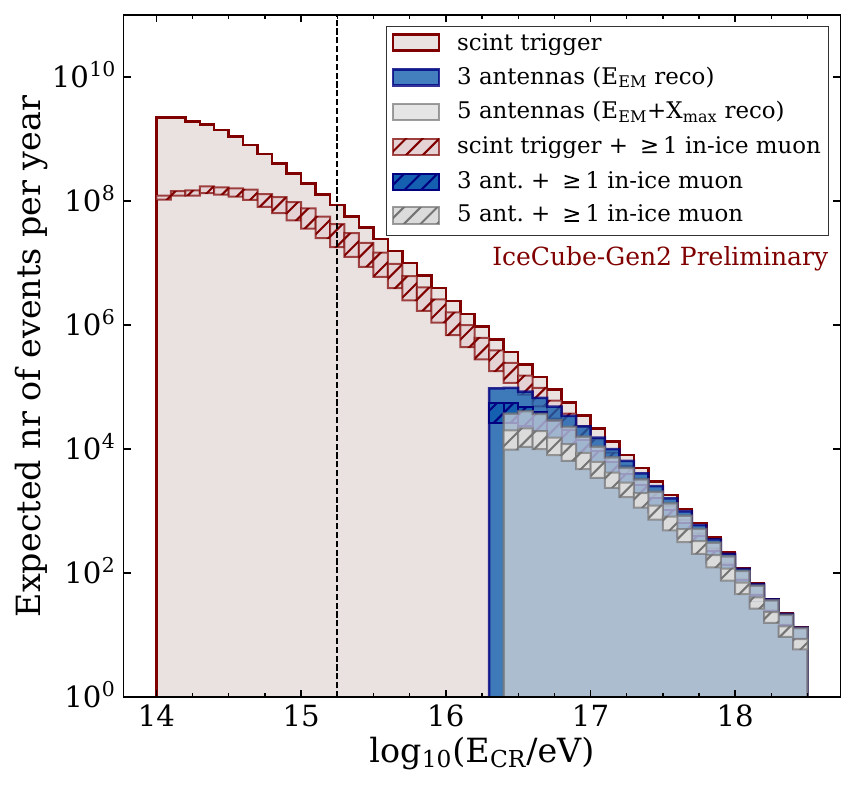}
  \hfill
    \includegraphics[width=0.54\textwidth]{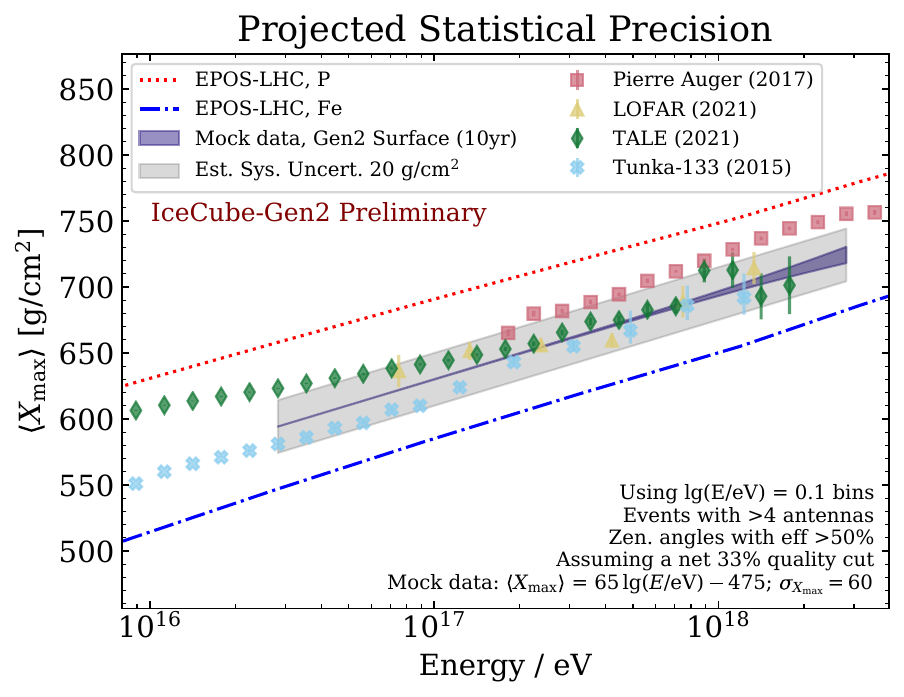}
  \shrink
  \end{center}
  \caption{Left: Expected event rate in lg(E) = 0.1 energy intervals for several event classes. The red distributions show the rates with scintillator information, while the blue ones are for events with radio information. The hashed regions indicate the range of events that are expected to have in-ice muon information as well. The vertical dashed line indicates where the scintillators have a $>$\,98\% reconstruction efficiency for contained events. Based on the work from~\cite{IceCube-Gen2:2021jce}. Right: The projected statistical and systematic resolution on $X_{\rm max}$ is shown for an assumed true distribution. The measurements from other experiments are shown with statistical error bars~\cite{Bellido:2017cgf,Corstanje:2021kik,TelescopeArray:2020bfv,Prosin:2015voa}.}
    \label{fig:xmax_and_count}
\end{figure}
The key to making further progress in the field of cosmic-ray physics is a better classification of the primary mass. Since simulations are necessary to interpret the mass of cosmic-ray observations using air shower arrays, there is an inextricable link between the quality of our understanding of particle physics and our understanding of astrophysics. Models of the hadronic production in air showers have been shown to have disagreements with observations across a broad range of energies~\cite{Soldin:2021wyv}. It is then critical to have an improved understanding of hadronic physics beyond the phase space of the LHC~\cite{Coleman:2022abf}. The surface array of IceCube-Gen2 will play an important role in making progress towards both improved measurements of mass and a better understanding of hadronic physics.

Already IceCube plays a fundamental role in providing constraints to model builders by mapping out the muon production at two largely separated energies, 1\,GeV~\cite{IceCubeCollaboration:2022tla} and 500\,GeV~\cite{verpoest:2023icrc}, constraining the phase space of potential enhancements to the models. For IceCube-Gen2, the measurements will be improved in two ways. Firstly, a larger range of zenith angles can be probed, implicitly studying the kaon/pion ratio early in the shower development. Secondly, the antennas will provide a pure measurement of the electromagnetic component of air showers. This will provide a calorimetric measurement of the air shower energy, reducing systematic uncertainties on the energy scale.

A related and important science case for the surface array is the study of prompt muons. For studying neutrinos, this is a primary background for when studying sources in the northern hemisphere. However, since these particles are produced by the high-energy particles within an air shower, they map out the early stages of development when the stochastic processes have not yet smeared out information from the first interaction. The decreased energy threshold provided by the scintillator array will enable reconstructions of energies on the surface and in the ice simultaneously, even for showers that land outside of the surface array~\cite{leszczynska:2023icrc}.

The energy measurements from the antennas, scintillator panels, and/or IceTop tanks, can be combined with measurements of \xmax~\cite{turcotte:2023icrc}. This quantity has been an industry-standard method for measuring the composition of primary particles using the electromagnetic cascade. The expected statistical resolution on \xmax using radio antennas for IceCube-Gen2 is given in the right panel of~\cref{fig:xmax_and_count}, assuming that $\leq$\,20\,g\,cm$^2$ accuracy can be achieved using five antennas, as has been shown for other radio-based measurements~\cite{Buitink:2014eqa,Bezyazeekov:2018yjw}. The IceAct telescopes will also cover the low-energy region, starting above 50\,TeV, again with sensitivity to both the electromagnetic content and \xmax~\cite{paul:2023icrc}.

\begin{figure}[tb]
  \shrink
  \begin{center}
    \includegraphics[width=0.48\textwidth]{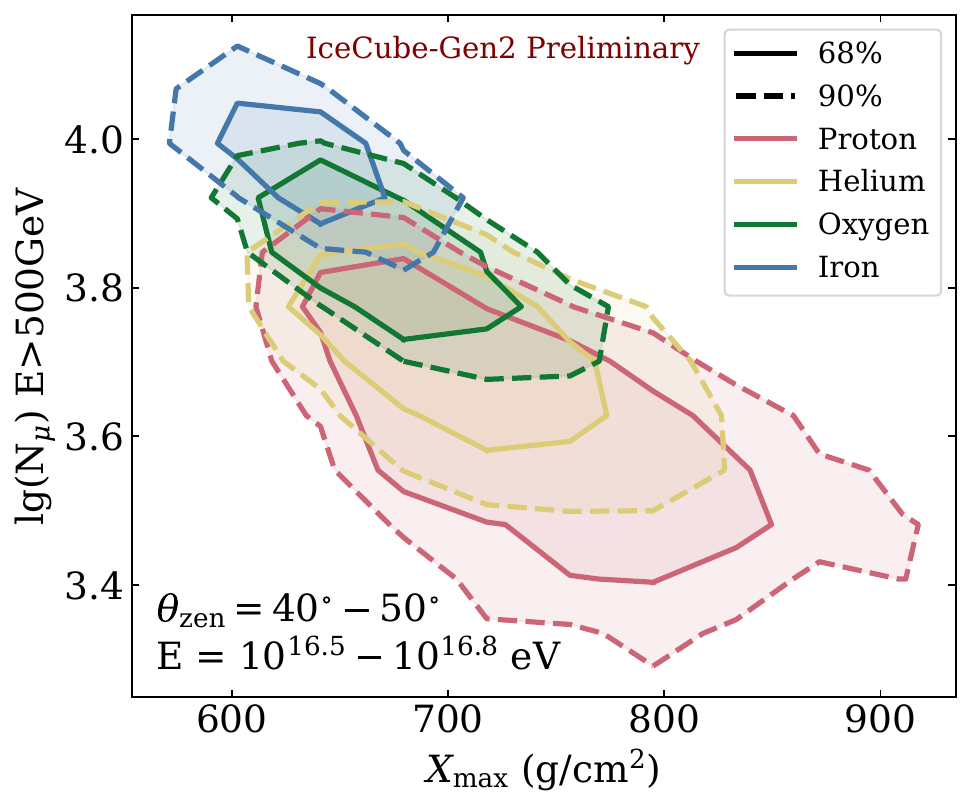}
    \hfill
    \includegraphics[width=0.505\textwidth]{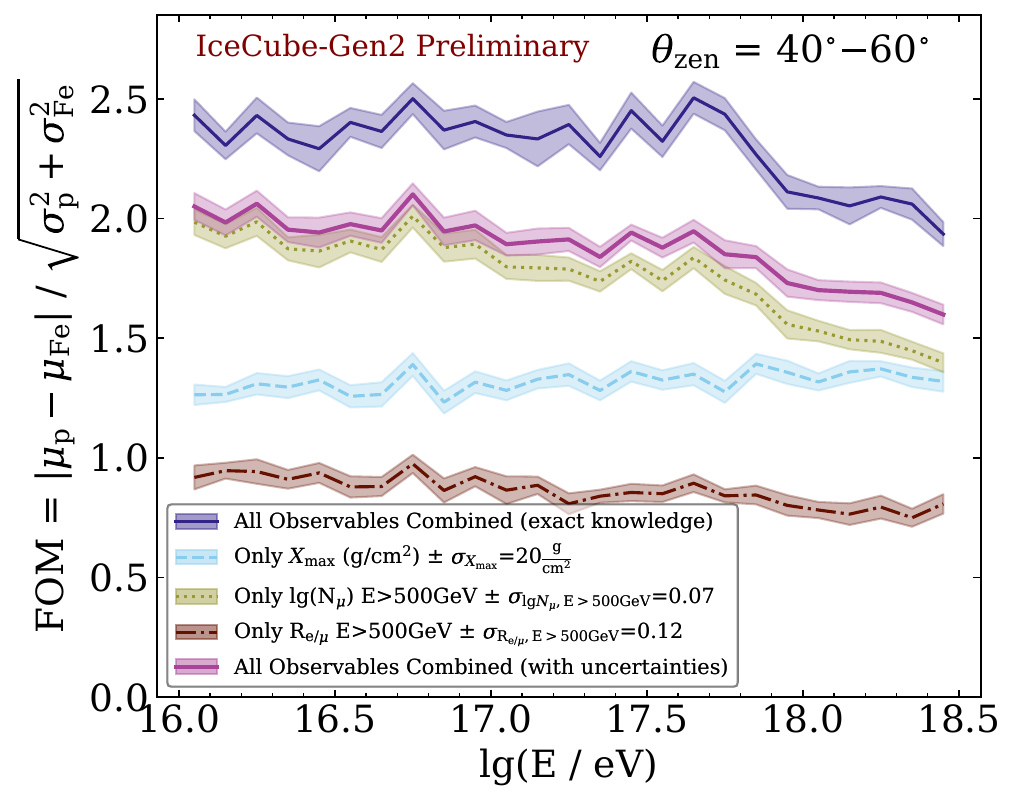}
  \end{center}
  \shrink
  \caption{Left: Contours of the true distributions of $X_{\rm max}$ and the number of high energy muons in air showers for four cosmic ray masses. The 68\% and 90\% resolutions are indicated by the solid and dashed lines. The number of high-energy muons has been scaled to account for the Heitler-Matthews beta exponent, see Ref.~\cite{Matthews:2005sd}. Right: A figure-of-merit for the separation power of proton and iron using various cosmic ray observables. The value for exact knowledge is shown as well the value of individual observables with expected resolutions, see Ref.~\cite{flaggs@arxiv}.}
    \label{fig:mass}
\end{figure}

Combining the observations of all the IceCube-Gen2 detectors will be important for mass separation. The left panel of~\cref{fig:mass} highlights the importance of making simultaneous measurements of high energy muons and \xmax. The distributions projected in this plane is shown for various mass groups.
The quantitative separation power of various air shower observables are shown in the right panel of~\cref{fig:mass} for the South Pole. The figure of merit (FOM) estimates the separation of proton and iron distributions in terms of number of sigmas\footnote{Specifically, this is the sum of the quadratically summed variances of distributions $A$ and $B$, $\sigma_{\rm total} = \sqrt{\sigma^2_A + \sigma^2_B}$.}, using a Fisher discriminant analysis. The FOM is shown for individual observables, spread by expected resolutions, such as those listed above, as well as the number of low energy muons~\cite{weyrauch:2023icrc} and the ratio of the muons and electromagnetic particles. The combined information, with expected resolutions, is shown in pink. This estimates that proton and iron can be separated by about two sigma. Additionally, the most influential observable is that of the high-energy muons ($>$\,500\,GeV) further highlighting the importance of the multi-observation design of IceCube(-Gen2). The separation from this observable decreases at higher energies due to the reduced separation of the proton/iron distributions. At the highest energies, \xmax and the number of high-energy muons become equally important. Further details on this study can be found in Ref.~\cite{flaggs@arxiv}.

\section{Improvement for neutrino astronomy}

\begin{figure}[tb]
  \shrink
  \begin{center}
    \includegraphics[width=0.99\textwidth]{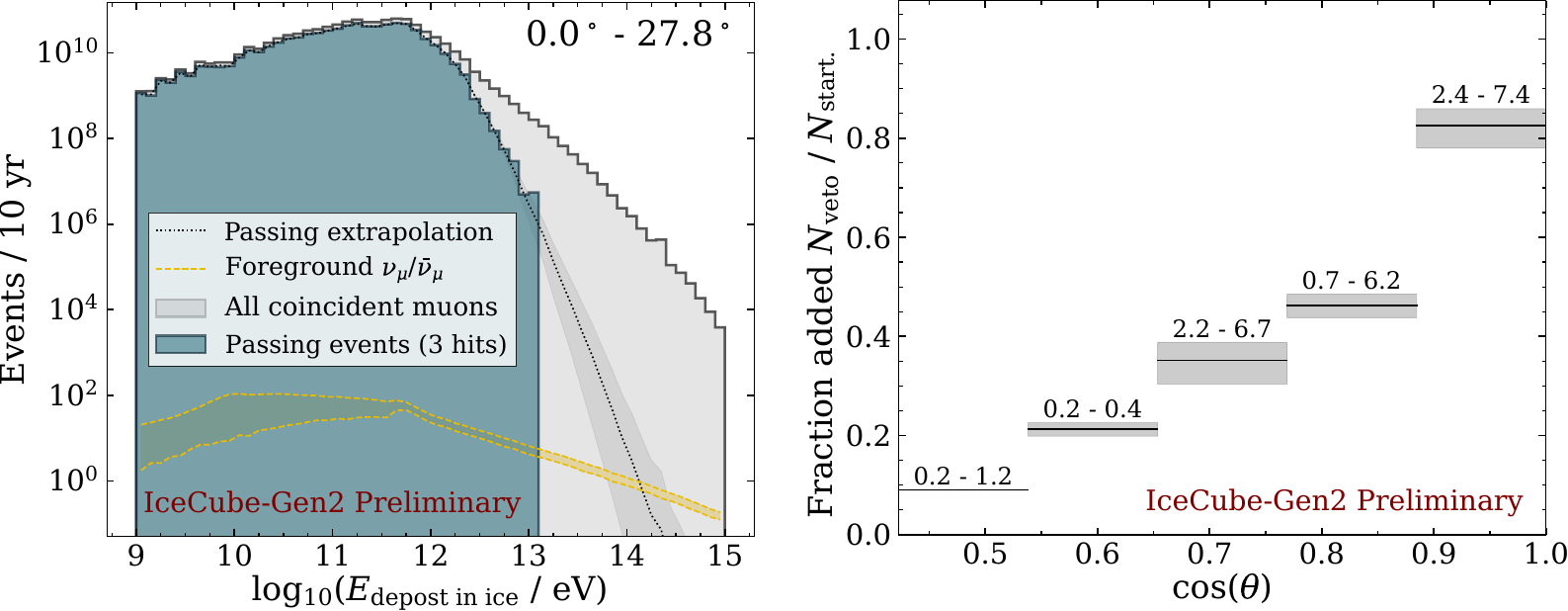}
  \end{center}
  \shrink
  \caption{Left: The total number of expected muons that pass through both the surface array and in-ice volume is shown as a function of deposited energy in the optical volume. The subset of those events that would not be rejected by a surface veto are shown in aqua. An extrapolation and the 1$\sigma$ statistical error of this passing fraction is shown by the back line and gray band. The foreground of muon neutrinos with coincident trajectories is shown in yellow. Right: The fractional increase in the total number of muon neutrinos that can be identified in the northern sky due to the veto is shown as a function of zenith angle. The number of events in each bin that would be added over ten years is indicated.}
    \label{fig:veto}
\end{figure}
An important consideration for the design of the surface array is to improve the neutrino-astronomy science goals of the observatory. While the background from cosmic rays make studying sources in the northern sky difficult, the surface array can reduce the background by acting as an air-shower veto. The layout of the stations enables sensitivity to 0.5\,PeV air showers with a requirement of 3 panels being triggered. The left panel of~\cref{fig:veto} shows the expected flux of muons and muon-neutrinos with a trajectory that pass through the surface array and deposit energy in the instrumented in-ice volume. The flux is calculated using the Sibyll~2.3d hadronic interaction model~\cite{Riehn:2019jet} and H4a as a flux/composition model~\cite{Gaisser:2011klf}. The neutrino flux is calculated using the diffuse flux measurements by IceCube~\cite{IceCube:2021uhz}.

It is clear that a reduction of five to eight orders of magnitude is required to reveal the neutrino flux. The surface can veto muons with $\sim$\,10\,TeV of deposited energy (the typical proxy for neutrino energy) and at $\sim$\,100\,TeV, the measurement is background-free. The right panel of~\cref{fig:veto} shows the relative increase in the neutrino rate for different zenith angle ranges. The 68\% confidence interval of the total number of extra neutrinos that would be observed from the diffuse flux over ten years of operation, is shown above each line. In total, the using the surface array a veto would improve the total number of astrophysical neutrinos in the southern sky by $11.0^{\,+6.2}_{-3.3}$ over ten years of operation.
\section{Summary}

The surface array of IceCube-Gen2 will play an important role in furthering our understanding of astroparticle physics. This will come about from two sources, the increased aperture of the array by a factor of about 30 and the combined detection of air showers using scintillator panels, radio antennas, water-Cherenkov tanks, and in-ice measurements.  

The increase in total surface array will generally enhance many of the existing cosmic-ray studies from IceCube/IceTop. For measurements of cosmic-ray anisotropy~\cite{mcnally:2023icrc,hou:2023icrc} and the all-particle and mass-dependent energy spectra~\cite{IceCube:2019hmk}, the increased statistics will allow for an extension into the $>\,10^{18}$\,eV energy range. Measurements of low energy ($\sim\,1$\,GeV) muons, where the main sensitivity comes from the tails of the shower footprint, will be directly benefited from the expanded size.

The additional detector components will allow for an expanded set of mass-dependent observables. The inclusion of \xmax~\cite{turcotte:2023icrc} and total number of muons~\cite{weyrauch:2023icrc} on a per-event basis will be a powerful tool to study the sources and propagational effects. In particular, these can be used to compare with other experiments where these two variables have been shown to disagree at high energies~\cite{Soldin:2021wyv}. Extending the IceCube measurements up to $\geq$\,EeV energies will allow us to study where this so-called muon puzzle begins, using a single detector.

Ultimately, a better understanding of hadronic interactions at high energies is a major goal in several subfields of astroparticle physics. The expanded aperture for coincident detection will be important for making tests of high-energy muon production. While the production of 500\,GeV muon production is currently being studied for vertical ($<$\,$35^\circ$) showers, for IceCube-Gen2, more inclined showers can be used, eliminating portions of the parameter space where solutions to e.g. the muon puzzle may have otherwise degenerate solutions~\cite{Riehn:2019jet}. Such an improvement in our understanding of hadronic physics would be important to all existing and previous experiments.

The surface array will also boost the primary focus of IceCube-Gen2, searching for $\geq$\,1PeV neutrinos. While on one hand, an improved knowledge of the processes and origins of the primary source of background across the whole sky will improve systematic uncertainties, the surface array will provide a veto and directly enhance the number of events in the southern sky. 

In total, the importance of the array extends into several areas of particle and astroparticle physics. The important role that IceCube plays currently in providing answers to where the highest energy particles are produced and how they arrive at Earth will be enhanced greatly by the IceCube-Gen2 surface array.

\bibliographystyle{ICRC}
\setlength{\bibsep}{1.pt}
\bibliography{references}

\clearpage

\section*{Full Author List: IceCube-Gen2 Collaboration}

\scriptsize
\noindent
R. Abbasi$^{17}$,
M. Ackermann$^{76}$,
J. Adams$^{22}$,
S. K. Agarwalla$^{47,\: 77}$,
J. A. Aguilar$^{12}$,
M. Ahlers$^{26}$,
J.M. Alameddine$^{27}$,
N. M. Amin$^{53}$,
K. Andeen$^{50}$,
G. Anton$^{30}$
% ...
\\
\\
% ...
$^{12}$ Universit{\'e} Libre de Bruxelles, Science Faculty CP230, B-1050 Brussels, Belgium \\
% ...
$^{17}$ Department of Physics, Loyola University Chicago, Chicago, IL 60660, USA \\
% ...
$^{22}$ Dept. of Physics and Astronomy, University of Canterbury, Private Bag 4800, Christchurch, New Zealand \\
% ...
$^{26}$ Niels Bohr Institute, University of Copenhagen, DK-2100 Copenhagen, Denmark \\
$^{27}$ Dept. of Physics, TU Dortmund University, D-44221 Dortmund, Germany \\
% ...
$^{30}$ Erlangen Centre for Astroparticle Physics, Friedrich-Alexander-Universit{\"a}t Erlangen-N{\"u}rnberg, D-91058 Erlangen, Germany \\
% ...
$^{47}$ Dept. of Physics and Wisconsin IceCube Particle Astrophysics Center, University of Wisconsin{\textendash}Madison, Madison, WI 53706, USA \\
% ...
$^{50}$ Department of Physics, Marquette University, Milwaukee, WI, 53201, USA \\
% ...
$^{53}$ Bartol Research Institute and Dept. of Physics and Astronomy, University of Delaware, Newark, DE 19716, USA \\
% ...
$^{76}$ Deutsches Elektronen-Synchrotron DESY, Platanenallee 6, 15738 Zeuthen, Germany  \\
$^{77}$ Institute of Physics, Sachivalaya Marg, Sainik School Post, Bhubaneswar 751005, India \\
% ...

\subsection*{Acknowledgements}

\noindent
The authors gratefully acknowledge the support from the following agencies and institutions:
USA {\textendash} U.S. National Science Foundation-Office of Polar Programs,
U.S. National Science Foundation-Physics Division,
U.S. National Science Foundation-EPSCoR,
Wisconsin Alumni Research Foundation,
Center for High Throughput Computing (CHTC) at the University of Wisconsin{\textendash}Madison,
Open Science Grid (OSG),
Advanced Cyberinfrastructure Coordination Ecosystem: Services {\&} Support (ACCESS),
Frontera computing project at the Texas Advanced Computing Center,
U.S. Department of Energy-National Energy Research Scientific Computing Center,
Particle astrophysics research computing center at the University of Maryland,
Institute for Cyber-Enabled Research at Michigan State University,
and Astroparticle physics computational facility at Marquette University;
Belgium {\textendash} Funds for Scientific Research (FRS-FNRS and FWO),
FWO Odysseus and Big Science programmes,
and Belgian Federal Science Policy Office (Belspo);
Germany {\textendash} Bundesministerium f{\"u}r Bildung und Forschung (BMBF),
Deutsche Forschungsgemeinschaft (DFG),
Helmholtz Alliance for Astroparticle Physics (HAP),
Initiative and Networking Fund of the Helmholtz Association,
Deutsches Elektronen Synchrotron (DESY),
and High Performance Computing cluster of the RWTH Aachen;
Sweden {\textendash} Swedish Research Council,
Swedish Polar Research Secretariat,
Swedish National Infrastructure for Computing (SNIC),
and Knut and Alice Wallenberg Foundation;
European Union {\textendash} EGI Advanced Computing for research;
Australia {\textendash} Australian Research Council;
Canada {\textendash} Natural Sciences and Engineering Research Council of Canada,
Calcul Qu{\'e}bec, Compute Ontario, Canada Foundation for Innovation, WestGrid, and Compute Canada;
Denmark {\textendash} Villum Fonden, Carlsberg Foundation, and European Commission;
New Zealand {\textendash} Marsden Fund;
Japan {\textendash} Japan Society for Promotion of Science (JSPS)
and Institute for Global Prominent Research (IGPR) of Chiba University;
Korea {\textendash} National Research Foundation of Korea (NRF);
Switzerland {\textendash} Swiss National Science Foundation (SNSF);
United Kingdom {\textendash} Department of Physics, University of Oxford.

\end{document}